\documentclass[aps,prl,twocolumn,reprint]{revtex4-1}
\usepackage{amsmath,amssymb,graphicx,hyperref,color}
\newcommand{\eqn}[1]{\begin{eqnarray} #1 \end{eqnarray}}
\newcommand {\ket}[1] {| #1 \rangle}
\newcommand {\bra}[1] {\langle #1 |}
\newcommand {\tbf}[1] {\textbf{#1}}
\newcommand {\tit}[1] {\textit{#1}}
\newcommand {\trm}[1] {\textrm{#1}}

\begin{document}

\title{A separable, dynamically local ontological model of quantum mechanics}

\author{Jacques Pienaar}
 \email{jacques.pienaar@univie.ac.at}
 \affiliation{
 Faculty of Physics, University of Vienna, Boltzmanngasse 5, A-1090 Vienna, Austria.
}
 \affiliation{
 Institute of Quantum Optics and Quantum Information,
Austrian Academy of Sciences, Boltzmanngasse 3, A-1090 Vienna, Austria.
}

\begin{abstract}
{A model of reality is called separable if the state of a composite system is equal to the union of the states of its parts, located in different regions of space. Spekkens has argued that it is trivial to reproduce the predictions of quantum mechanics using a separable ontological model, provided one allows for arbitrary violations of `dynamical locality'. However, since dynamical locality is strictly weaker than local causality, this leaves open the question of whether an ontological model for quantum mechanics can be both separable and dynamically local. We answer this question in the affirmative, using an ontological model based on previous work by Deutsch and Hayden. Although the original formulation of the model avoids Bell's theorem by denying that measurements result in single, definite outcomes, we show that the model can alternatively be cast in the framework of ontological models, where Bell's theorem does apply. We find that the resulting model violates local causality, but satisfies both separability and dynamical locality, making it a candidate for the `most local' ontological model of quantum mechanics.}
\end{abstract}

\maketitle

\section{Introduction}

``We are all made of algebra-stuff: the elements of local reality are faithfully described not by real variables or stochastic real variables but by the elements of a certain algebra that can be represented by Hermitian matrices. [...] A failure to understand this can result in various misconceptions about quantum physics in general, and about its locality in particular. " \\ 
-- David Deutsch \cite{DEUVIN} \\

Can we explain quantum phenomena in terms of an underlying model of reality? Traditionally, such explanations took the form of hidden variable models. Bell's theorem places a severe constraint on these models, demonstrating that (given certain reasonable assumptions) no hidden variable model can reproduce the predictions of quantum mechanics and at the same time satisfy \tit{local causality} \cite{BELL76, WISC}.

Since Bell's work, hidden variable models have been generalised and extended to the framework of \tit{ontological models}, introduced by Harrigan \& Spekkens \cite{HAR, SPE05}. While the framework is not general enough to capture all possible interpretations of quantum mechanics, it is sufficiently general to permit a clear exposition of the various definitions of locality that are central to understanding the implications of Bell's theorem. 

We will focus on ontological models set in relativistic space-time, in which one can ascribe a real, physical state, called the \tit{ontic state}, to the physical systems located in any spatial region at one time (more generally, on a space-like hyper-surface). Given a foliation of space-time into a family of time-slices (or hyper-surfaces), the ontic state evolves with respect to the global time parameter according to some dynamical laws. Measurements performed in a region of space-time produce outcomes with probabilities that depend only on the ontic state in the region where the measurement is performed. We do not make any assumptions about the kinematics and evolution of the ontic states, other than that they reproduce the predictions of quantum mechanics. For example, quantum mechanics satisfies \tit{signal locality}, which forbids sending information faster than light, so this should be upheld by the predictions of the ontological model. However, this does not prevent the ontic states from moving faster than light, as occurs for example in the De-Broglie Bohm model\cite{BOHM}.

Einstein, in correspondence with Schr\"{o}dinger, mentioned two different conceptions of locality in physics \cite{HOW85, HAR}. First, the notion of \tit{separability}, which says that the joint ontic state of the physical systems in two regions of space at a given time is just the union of the states in the sub-regions at that time (``the real state of $(AB)$ consists precisely of the real state of $A$ and the real state of $B$, which two states have nothing to do with one another"). Second, \tit{dynamical locality}, which implies that changes made to one system should not be able to influence the state of another system at space-like separation from the first (``The real state of $B$ thus cannot depend upon the kind of measurement I carry out on $A$"). The first can be read as an essentially kinematical statement about how states compose to form composite states, while the second can be read as a dynamical statement about how the global ontic state changes in response to a local measurement. We will provide more precise definitions of these concepts later on.

The union of \tit{separability} and \tit{dynamical locality} are not enough to imply Bell's notion of \tit{local causality}. For that, we need the additional property of \tit{factorisation}, which, roughly speaking, requires the probabilities of space-like separated measurement outcomes to be independent conditional on the joint ontic state just prior to measurement. While the first two notions of locality fit nicely with the world-view expounded by classical relativistic physics, the factorisation property is a more mysterious and less obvious pre-requisite for a local relativistic theory. We will discuss its physical meaning and justification in the Discussion. It turns out that (given the basic assumptions of the framework of ontological models) \tit{dynamical locality} and \tit{factorisation} alone are sufficient for \tit{local causality}. As emphasised by Henson, \tit{separability} plays no part in this implication -- a theory may be locally causal with or without being separable \cite{HEN}. 

Spekkens has argued that it is trivial to construct ontological models of quantum mechanics that satisfy \tit{separability}, provided one abandons \tit{dynamical locality} \cite{SPEX}. According to Spekkens, this points to a fundamental absurdity in treating \tit{separability} and \tit{dynamical locality} as distinct concepts. Instead, the false distinction between kinematics and dynamics should be erased and replaced by a unifying framework, such as causal structure. If we accept this idea, then a reading of Bell's theorem that implies a failure of \tit{dynamical locality} ought to suggest a failure of \tit{separability} as well (as occurs, for example, in the De-Broglie Bohm model). On the other hand, Wiseman and Cavalcanti have pointed out that one can keep a notion of \tit{dynamical locality} if one is willing to reject \tit{factorisation}\footnote{To be precise, the authors show that one can keep `Relativistic Causality' and `Common Causes' at the expense of `Decorrelating Explanation'. The first two can be shown to imply \tit{dynamical locality} and we interpret the latter as equivalent to \tit{factorisation}.}\cite{WISC}. But can we not also retain \tit{separability}?

The literature would suggest not: all known ontological models that satisfy \tit{dynamical locality} (or something similar) reject \tit{separability}. Why should this be the case? Most likely, it is because all such models take the wave-function or density matrix to represent all or part of reality. In the words of Spekkens, ``It turns out that for any realist interpretations of quantum theory wherein the ontic state encodes the quantum state [...] kinematical locality fails simply by virtue of the existence of entangled states" \cite{SPEX}. However, there exist formulations of quantum mechanics in which the dynamical objects are not wave-functions but algebraic operators, the best-known example being relativistic quantum field theory, where field operators are the main candidates for the elements of reality. We need not go so far as field theory: Deutsch and Hayden \cite{DH}, following Gottesman \cite{GOT}, reformulated non-relativistic quantum mechanics in terms of local operators evolving in the Heisenberg picture, and used their model to argue that quantum mechanics is both separable and local.

Timpson\cite{TIM03} has argued that Deutsch and Hayden's model inherits its (dynamical) locality from the many-worlds interpretation, in particular because this interpretation allows it to evade Bell's theorem. Timpson emphasised that the model's primary feature is not dynamical locality, but rather \tit{separability}, which allows composite states to be reduced to their subsystems\cite{TIM03,WALT14}. Here, we re-cast the model of Deutsch and Hayden in the framework of ontological models, which does not assume the many-worlds interpretation and is therefore vulnerable to Bell's theorem. Taking the operators of Deutsch and Hayden as the ontic states, we show that the resulting ontological model still satisfies both \tit{separability} and \tit{dynamical locality}, though it fails \tit{factorisation}. We argue that this is the strongest possible manifestation of locality in an ontological model that can reproduce quantum predictions. We discuss the interesting features of the model, as well as its drawbacks.

\section{Background}

We work in the framework of \tit{ontological models} \cite{HAR,SPE05}. This assumes, among other things, that experimenters are `free' \footnote{Specifically, we assume that the causes of the experimenter's choice are not relevant to the variables under investigation. Our assumption is equivalent to `Free Choice' as described in Ref. \cite{WISC}.} to choose their measurement settings and that a given measurement outcome is represented for all observers by a single stochastic variable. It therefore implicitly excludes interpretations of quantum mechanics that avoid Bell's theorem by appealing to eg. super-determinism or observer-relativism as occurs in some versions of the many-worlds interpretation. \\

\tbf{Definition:} An \tit{ontological model} is a prescription for the elements of reality that aims to tell us: (i) what the `real, physical states' that we prepare in the laboratory are, (ii) how these states transform with time, and (iii) how they determine the probabilities of outcomes for a given measurement procedure. In particular,

(i) The model assumes that every system has a `real, physical state' $\lambda$, called the \tit{ontic state}, belonging to a set of possible ontic states $\Lambda$. For a given preparation procedure $P$ of the system, the model prescribes a probability distribution $\mu_P(\lambda)$ over the system's ontic states. If it is not a delta function, then $\mu_P(\lambda)$ is called an \tit{epistemic state} and it represents our subjective ignorance (given the preparation $P$) about which ontic state was actually prepared.

(ii) Letting the system evolve for a fixed time interval under controlled laboratory conditions is called a transformation $T$ of the system. Given a laboratory procedure $T$ for transforming the system, the ontological model prescribes a probability density $\Gamma_T(\lambda,\lambda')$ for the ontic state to change from $\lambda$ to $\lambda'$. If this function only takes values 0 or 1, the transformation is called \tit{deterministic}.

(iii) For any measurement procedure $M$ performed on the system, the ontological model prescribes a probability $\xi_{M,m}(\lambda)$ of obtaining the the outcome `$m$' when the system is in the ontic state $\lambda$. In the special case where this probability is always either 0 or 1, the ontological model is said to satisfy \tit{predetermination}, since knowing the ontic state is then sufficient to determine the outcome of any measurement. Note that a measurement does not affect the ontic state (although the epistemic probability can be updated conditional on the measurement outcome).

The above three ingredients combine to predict the probability of obtaining an outcome $m$, conditional on the preparation, transformation and measurement ($P,T,M$) performed on a physical system:
\eqn{
p(m|P,T,M) &=& \int \int d\lambda d\lambda' \, \mu(\lambda) \, \Gamma_T(\lambda', \lambda) \, \xi_{M,m}(\lambda') \, . \nonumber \\ 
&&
}
In what follows, we will only consider ontological models whose predictions agree with the predictions of quantum mechanics.\\

\begin{figure}[!htbp]
\includegraphics[width=8cm]{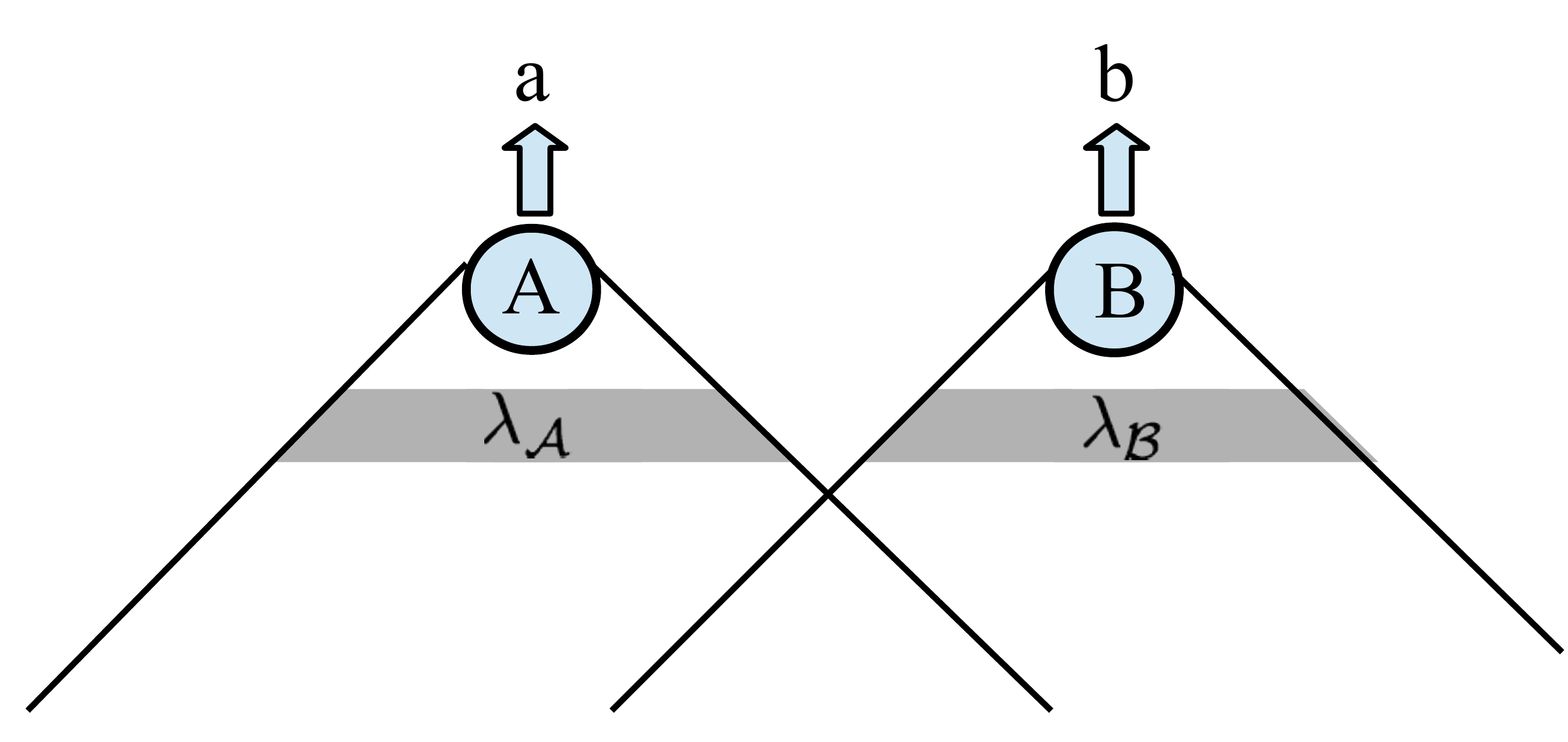}
\caption{A space-time schematic of independent operations $A,B$ performed on the ontic states in regions $\mathcal{A},\mathcal{B}$ and resulting in outcomes $a,b$ respectively. Time is on the vertical axis, space is the horizontal axis, and the diagonal lines denote the past light-cones of the operations. The shaded regions contain the ontic states just prior to the operations.}
\label{fig:lcone}
\end{figure}

As mentioned in the introduction, we consider ontological models embedded in space-time, so that we can assign an ontic state to the systems in a region of space at a given time. This allows us to discuss various concepts of locality. In the following definitions, $\mathcal{A}$ and $\mathcal{B}$ refer to disjoint regions of space (i.e. space-like separated in a relativistic setting), $a,b$ refer to the outcomes of measurements performed in those regions, and $A,B$ refer to local operations (including the choosing of measurement settings) performed in their respective regions, as depicted in Fig. \ref{fig:lcone}. In the definitions below, $A$ labels a particular choice of transformation and measurement procedure in region $\mathcal{A}$ represented by the transition matrix $\Gamma_{A}(\lambda_{\mathcal{A}}, \lambda'_{\mathcal{A}})$ and indicator function $\xi_{A,a}(\lambda_{\mathcal{A}})$ respectively, and a similar representation holds for the operations labelled by $B$. \\

\tbf{Definition: (S)} An ontological model satisfies \tit{separability} iff the state of systems in an extended region of space is equivalent to the union of the states of the systems in localised sub-regions. Formally, let $R$ be a region of space, partitioned into $N$ disjoint sub-regions $R_1,R_2,...R_N$. Let $\Lambda_{R_i}$ denote the ontic state space of systems in region $R_i$ at a given instant. Separability says that the state space of the total system is the Cartesian product of the state spaces of the sub-systems:
\eqn{
\Lambda_{R} = \Lambda_{R_1} \times \Lambda_{R_2} \times ... \times \Lambda_{R_N} \, .
}

\tbf{Definition: (SL)} An ontological model satisfies \tit{signal locality} iff the probability of outcome $a$ in region $\mathcal{A}$ is independent of the parameters $B$ selected in region $\mathcal{B}$ (and symmetrically for the outcomes $b$). Formally, signal locality holds iff:
\eqn{
p(a|A,B) = p(a|A) \, , \nonumber \\
p(b|A,B) = p(b|B) \, .
}
The property SL must hold if the model is to agree with the predictions of quantum mechanics. It is worth comparing signal locality to the stronger conditions of dynamical locality and parameter independence described below. \\

\tbf{Definition: (DL)} An ontological model satisfies \tit{dynamical locality} iff the ontic state $\lambda_{\mathcal{A}}$ in a spatial region $\mathcal{A}$ is independent of operations performed in a disjoint spatial region $\mathcal{B}$. A similar statement holds for the ontic state $\lambda_{\mathcal{B}}$ with respect to what is done in region $\mathcal{A}$. Formally, dynamical locality holds iff, after the local operations in $\mathcal{A}$ and $\mathcal{B}$, the final state $\lambda'_{\mathcal{A}}$ is independent of the parameter $B$, and $\lambda'_{\mathcal{B}}$ is independent of the parameter $A$. This definition differs from the more standard notion of \tit{parameter independence}:\\

\tbf{Definition: (PI)} A model satisfies parameter independence iff the probabilities of outcomes $a$ in $\mathcal{A}$ are independent of the parameter $B$ (and symmetrically for $b$). Formally, parameter independence holds iff:
\eqn{
p(a|A,B,\lambda_{\mathcal{AB} }) = p(a|A,\lambda_{\mathcal{AB}}) \, , \nonumber \\
p(b|A,B,\lambda_{\mathcal{AB}}) = p(b|B,\lambda_{\mathcal{AB}}) \, .
}

Since measurement outcomes in a region can only depend on the ontic state in that region, dynamical locality implies parameter independence. The converse is not true: it is possible for the ontic state to be affected by a remote operation, without the probabilities being affected. Similarly, parameter independence implies signal locality, but the converse does not hold, so $DL \implies PI \implies SL$. While derivations of Bell's theorem typically favour the weakest possible assumptions, our goal here is to find the strongest manifestation of locality compatible with quantum mechanics, so we prefer DL to parameter independence. \\

\tbf{Definition: (F)} An ontological model satisfies \tit{factorisation} if the probabilities of measurement outcomes $a$ are independent of the measurement outcomes $b$, conditional on the settings $A,B$ and the joint ontic state $\lambda_{\mathcal{AB}}$ prior to measurement. Formally, factorisation holds iff: \\
\eqn{
p(a,b|A,B,\lambda_{\mathcal{AB}}) = p(a|A,B,\lambda_{\mathcal{AB}}) \, p(b|A,B,\lambda_{\mathcal{AB}}) \, . \nonumber \\
&& 
}

\tbf{Definition: (LC)} An ontological model satisfies \tit{local causality} iff the probabilities of measurement outcomes $a$ are independent of the settings $B$ and the outcomes $b$, conditional on the joint ontic state $\lambda_{\mathcal{AB}}$ prior to measurement. A similar statement holds for the outcomes $b$ with respect to the settings and outcomes $A,a$. Formally, local causality holds iff: \\
\eqn{
p(a|b,A,B,\lambda_{\mathcal{AB}}) = p(a|A,\lambda_{\mathcal{AB}}) \, , \nonumber \\
p(b|a,A,B,\lambda_{\mathcal{AB}}) = p(b|B,\lambda_{\mathcal{AB}}) \, .
}

Bell's theorem can now be stated as: \tit{no ontological model satisfying local causality can reproduce the predictions of quantum mechanics.} As a corollary, we note that DL \& F $\implies$ LC, so Bell's theorem rules out at least one of DL or F. We will see that the Deutsch-Hayden model provides an example in which both S and DL are satisfied, at the expense of F. If (as we argue in the Discussion) a failure of F does not by itself represent any form of non-locality, then it follows that this is the strongest possible manifestation of locality in an ontological model compatible with quantum mechanics.

\section{The Deutsch-Hayden ontological model}

The ontological model we describe here will be called the \tit{DH-model}, although it deviates from Deutsch and Hayden's original formalism (see eg. Refs. \cite{DH,DEUVIN,TIM03,HORS}). In particular, the authors avoided Bell's theorem by denying that measurement outcomes could be represented objectively by stochastic random variables: \\

``[T]he false premise [in Bell's theorem] occurs in the first sentence of the argument, where we assumed that we could assign stochastic variables [...] to the `actual outcomes' of measurements." \cite{DH} \\

By contrast, we aim to `take the Bell by the horns' and cast the formalism in a way that is vulnerable to Bell's dilemma. To do so, we recast the formalism in a manner compatible with the framework of ontological models. 

Before we begin, it will be useful to distinguish two types of abstract spaces ascribed to a physical system. On one hand, there is the usual Hilbert space of the system's wave-function; on the other hand, there is the space of ontological states. Since both might refer to the dimension of some Hilbert space, to avoid confusion we refer to the first as the $\psi$-\tit{space} and the second as the \tit{ontic state space}. In a $\psi$-\tit{complete} ontological model (see Ref. \cite{HAR}), these spaces are the same, because the ontic state is identified with the quantum state of the system. However, in general, as in the present model, the ontic state space is larger than the $\psi$-space. 

\tit{Ontic states---} Let us consider a universe composed of $N$ two-level systems (qubits). The ontic state of a qubit at time $t$ is a triple of operators $\lambda(t) := \{\bar{X}(t),\bar{Y}(t), \bar{Z}(t) \}$, where the bar indicates that the operator has the form of an operator on the single-qubit $\psi$-space, sandwiched by $N-1$ qubit identity operators, for example $\bar{X}(t) := I \otimes I \otimes ... X(t) ... \otimes I \otimes I $. The single-qubit operators $X(t),Y(t),Z(t)$ are conventionally represented by the Pauli matrices at some designated initial time $t=0$:
\eqn{
X = \left[ \begin{array} {ll}
0 & 1 \\ 
1 & 0  \\
\end{array}
\right] \, , \,\,
Y = \left[ \begin{array} {ll}
0 & -i \\ 
i & 0  \\
\end{array}
\right] \, , \,\,
Z = \left[ \begin{array} {ll}
1 & 0 \\ 
0 & -1  \\
\end{array}
\right] \, ,
}
and the $+1$ eigenstate of the $Z$ operator, $\ket{z_+}$, is represented by the triple $\{\bar{X},\bar{Y},\bar{Z}\}$. The ontic state of an arbitrary pure state $\ket{\psi}:= U_{\psi} \ket{z_+}$ is just given by $U^{\dagger}_{\psi} \{ \bar{X},\bar{Y},\bar{Z} \} U_{\psi}$, so for example,
\eqn{ 
\ket{z_{-}} &\leftrightarrow& \{ \bar{X},-\bar{Y},-\bar{Z} \}, \, \nonumber \\
\ket{x_{+}} &\leftrightarrow& \{ \bar{Z},-\bar{Y},\bar{X} \},  \, \nonumber \\
\trm{etc.} && 
}

Although each operator acts non-trivially on the $\psi$-space of just a single qubit, the padding by identity operators means that the barred operators $\bar{X}(t),\bar{Y}(t), \bar{Z}(t)$ act on the entire $\psi$-space of $N$ qubits. Thus, the ontic state space of a system in the DH-model is much larger than the $\psi$-space of the same system. The reason for this `excess baggage' will become clear shortly, when we discuss the dynamics of the model. 

We assume that an arbitrary physical system is composed of a finite integer number of qubits. For a system $S$ composed of $n$ qubits, the ontic state at time $t$ is given by the set of ontic states of its individual qubits, $\lambda_S(t) := \{ \lambda_i(t) : i=1,2,...,n \}$. 
Each of the three operators in the ontic state of a qubit can be represented by a $2^N \times 2^N$ matrix, or by a $2^{2N}$ dimensional vector. The triplet of these operators can therefore be represented by a $2^{6N}$ dimensional vector. If $\vec{\lambda_i}$ is the vector representing the ontic state of the $i_{\trm{th}}$ qubit, then the ontic state of $S$ is represented by the tensor product of its individual qubit vectors. Hence the ontic state space $\Lambda_S$ is isomorphic to the family of vectors in $2^{6Nn}$ dimensional Hilbert space that have the product form $\vec{\lambda_S} = \vec{\lambda_1} \otimes \vec{\lambda_2} \otimes ... \otimes \vec{\lambda_n}$. It follows that the ontic states of physical systems compose and decompose according to a Cartesian product rule: the ontic state space of two subsystems with state spaces $\Lambda_A$ and $\Lambda_B$ is just $\Lambda_A \times \Lambda_B$ (interpretable as the space of product vectors in the joint Hilbert space $\Lambda_A \otimes \Lambda_B$). It follows that the ontological model is separable.

To be more precise: we are assuming that every region of space contains a number of `physical' qubits whose Hilbert space encodes the degrees of freedom of all quantum matter that exists in that region. Then, separability follows from the composition of states in the model as described above. What are these `physical qubits' that form the building blocks of all systems? What is their density per unit of space-time? These questions are beyond the scope of this paper, but may be interesting to seek a connection with models of discretised space-time.

In the DH-model, mixed states can arise by a combination of two distinct processes: epistemic uncertainty, or entanglement to other systems. The first type of mixed state is represented by a distribution $\mu_P(\lambda_S)$ on the ontic state space that is not a delta functional. The second type is represented by a delta functional centred on an ontic state whose non-trivial elements (i.e. elements that are not the identity operator) act on a Hilbert space that is larger than the $\psi$-space of the system, for example, a qubit whose ontic state has an element of the form $I\otimes ... X \otimes Y ... \otimes I $, in which the part $ X \otimes Y$ acts non-trivially on the $\psi$-space of two qubits. In general, a preparation procedure can result in both types of mixing. The fact that the details of the preparation procedure can result in different ontic representations of the same mixed density matrix means that the model is \tit{preparation contextual} (see Ref. \cite{SPE05}).

\tit{Dynamics---} In the DH-model, arbitrary dynamics of a system is represented by applying the appropriate unitary operator (padded with identities as needed) to the ontic state. Thus, a qubit evolves as:
\eqn{
\lambda &=& \{ \bar{X},\bar{Y},\bar{Z} \} \nonumber \\
&\rightarrow& \bar{U}^{\dagger}(t) \{ \bar{X},\bar{Y},\bar{Z} \} \bar{U}(t) \nonumber \\
&=&  \{ \bar{X}(t),\bar{Y}(t),\bar{Z}(t) \} \nonumber \, \\
&:=& \lambda' \, .
}
Notice that, in the case where $\bar{U}(t)$ acts non-trivially on more than one qubit, we need to make use of the identities in the ontic state. For example, if $\bar{U}(t)$ represents a three-qubit operation, then the non-trivial part of $\bar{X}(t)$ looks like $U^{\dagger}(t) [\,X \otimes I \otimes I \,] U(t)$; for example, if after a time $t$ we have $U(t)=ZYZ$, then $X(t)=-XYZ$. Notice that the operator $\bar{X}(t)$ now acts non-trivially on a three-qubit subspace of the full ontic state space of that qubit. In general, evolutions that would normally be represented by a completely positive trace-preserving (CPTP) map in the standard formalism are represented in the DH-model by a unitary that acts on the system plus additional systems. Hence dynamics (prior to measurement) is assumed to always be unitary `in reality'. Unlike in the standard formalism, the choice of purification has a physical significance in the DH-model: the resulting ontic state of the system will depend on the choice of purification. In the above example, we only considered the dynamics of a single qubit, but the ontic state of this qubit spans a subspace of $\Lambda$ whose dimension depends on the number of qubits involved in the selected purification of the dynamics. Even if we never perform measurements on the other two qubits, the ontic state of the first qubit contains the information that exactly two (and not more) ancilla qubits played a non-trivial role in the transformation. In this sense, the model subscribes to a `church of the larger Hilbert space' in that the ancilla systems used in purifying the dynamics are assumed to be real, physical systems and not just a mathematical convenience. 

\begin{figure}[!htbp]
\includegraphics[width=8cm]{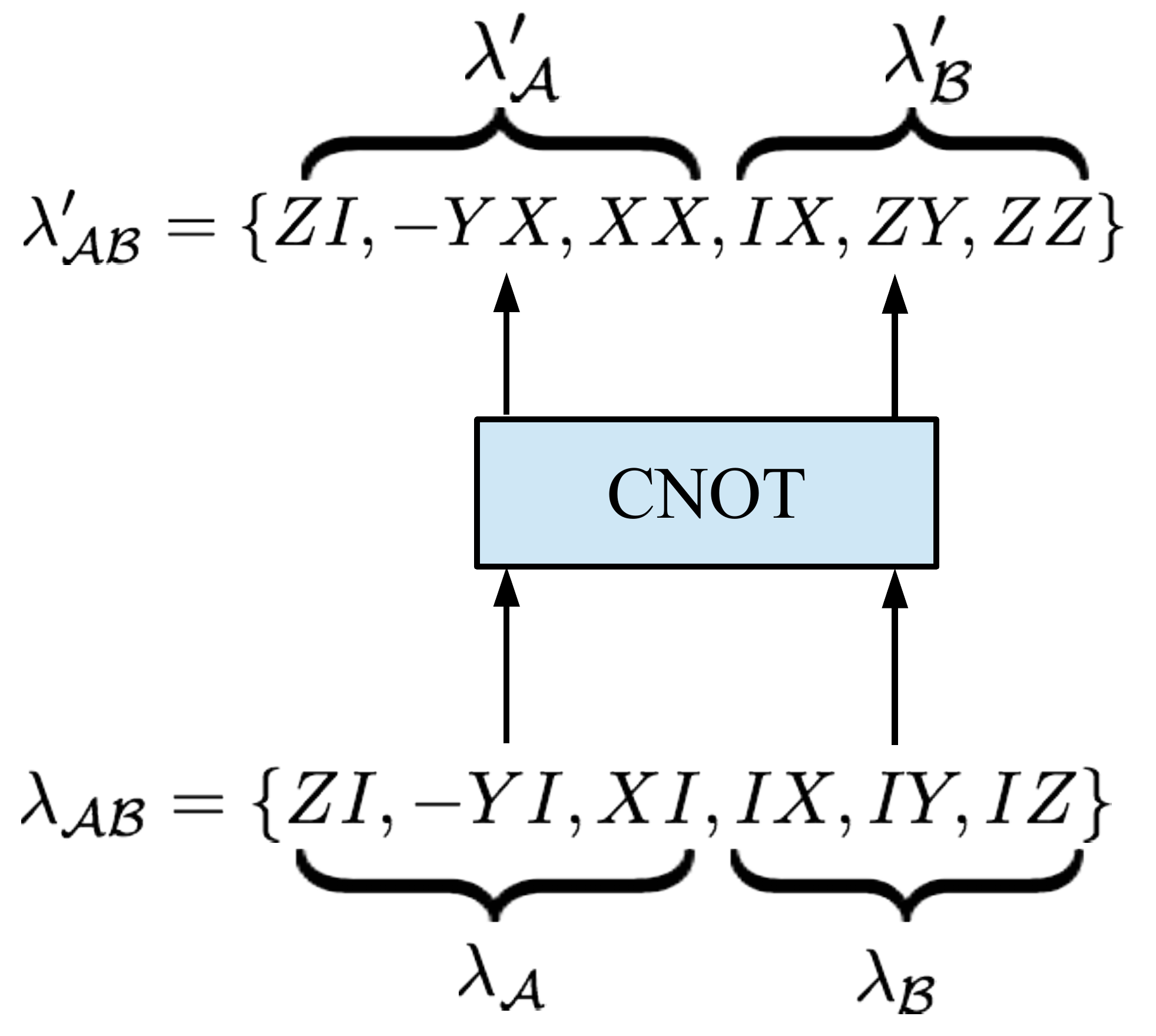}
\caption{A diagram showing the ontic states of two qubits before and after evolution through a CNOT gate, in a two-qubit `universe'. We employ the shorthand $XY:=X \otimes Y$. Here, $\lambda_{\mathcal{A}},\lambda_{\mathcal{B}}$ are the initial ontic states and $\lambda'_{\mathcal{A}},\lambda'_{\mathcal{B}}$ the final ontic states of the individual qubits. These represent an entangled state at the output, since the ontic states of each qubit contain terms acting non-trivially on the full two-qubit space. Nevertheless, the joint ontic state is separable.}
\label{fig:cnot}
\end{figure}

Fig. \ref{fig:cnot} shows the joint and individual ontic states of two qubits (assumed to be the only qubits in the universe) that are initially in a state corresponding to the product state $\ket{x_+}\ket{z_+}$, which become entangled by application of a quantum controlled-NOT gate (CNOT). The output $\{ \lambda'_{\mathcal{A}},\lambda'_{\mathcal{B}} \}$ corresponds to an entangled Bell state represented in the standard formalism by:
\eqn{
\ket{\psi}_{\mathcal{AB}} = \frac{1}{\sqrt{2}} (\ket{z_+}\ket{z_+}+\ket{z_{-}}\ket{z_{-}}) \nonumber \, . 
}

In general, a transformation $T$ of a system $S$ is represented by a transition matrix $\Gamma_T(\lambda'_S, \lambda_S)$ that specifies the probability of a unitary transformation $\lambda_S \rightarrow \lambda'_S := \bar{U}^{\dagger} \lambda_S \bar{U}$ for some unitary map $\bar{U}$ acting on a $2^{N}$ dimensional Hilbert space (Eg. for an $n$-qubit gate, the non-trivial part of $\bar{U}(t)$ acts on the $2^n$ dimensional subspace). As with preparations, the representation of a general transformation on a quantum system depends upon its purification to ancilla states, so the model is \tit{transformation contextual} in the language of Ref. \cite{SPE05}.

This explains why the dimensionality of the ontic state space needs to be so large: it has to be able to store information about all physical systems with which the system interacts during evolution. Intuitively, every system keeps a kind of `record' of its past local interactions with other systems, and since the size of this record can increase with time, the ontic state needs enough degrees of freedom to accommodate all possible histories of the system. This usage of `ontological excess baggage' bears some resemblance to Spekkens' separable (but dynamically nonlocal) model in which every local system carries a copy of the universal wavefunction \cite{SPEX}. The key difference is that in the DH-model, the information stored locally does not include the states of remote systems at the same time, which enables it to be both separable and dynamically local. It therefore should be possible to formulate the model in a way that does not rely on an artificial distinction between kinematics and dynamics, although this distinction is utilised for convenience in the present framework.

\tit{Measurements---} In standard quantum theory, a measurement $M$ on an $n$-qubit system resulting in an outcome $m$ is associated with a Hermitian operator (or POVM element) $O_{M,m}$ acting on the system's $2^n$ dimensional $\psi$-space. The set of these operators for all outcomes $m$ satisfies $\sum_m O_{M,m} = I$, with $I$ the identity on the $\psi$-space. Let $\bar{J}_i$ stand for any one of the operators $\bar{I},\bar{X}_i,\bar{Y}_i,\bar{Z}_i$ from the ontic state of the $i_{\trm{th}}$ qubit in $S$ (where $\bar{I}$ is just the identity on all $N$ qubits). Any operator $O_{M,m}$ can be padded with identities to cover a $2^N$ dimensional space (designated as before by a bar on the operator). The resulting operator $\bar{O}_{M,m}$ can then be expressed as a sum of up to $4^n$ terms, each of which is a product of $n$ operators $\bar{J}_i$:

\eqn{
\bar{O}_{M,m} &=& \sum^{4^n}_{j}  \,  \alpha_{M,m}(j) \, \prod^{n}_{i=1} \, \bar{J}_i \,  \nonumber \\
&:=& \mathcal{O}_{M,m}[\bar{J}_i ] \, ,
}
where the coefficients $\alpha_{M,m}(i,j)$ can be zero for some values of $j$. Since the non-trivial $\bar{J}_i$ are elements of the ontic state $\lambda_S$, this defines a function $\mathcal{O}_{M,m}[\lambda_S ]$, characterised by the $4^n$ coefficients $\alpha_{M,m}(j)$, which maps the ontic state to the operator $\bar{O}_{M,m}$ at time $t=0$. More generally, $\mathcal{O}_{M,m}$ maps the time-evolved ontic state $\lambda_S(t) = U^{\dagger}(t) \lambda_S  U(t)$ to the time-evolved operator $U^{\dagger}(t) \bar{O}_{M,k} U(t)$. Thus, the measurement $M$ is associated with a fixed function $\mathcal{O}_{M,m}[...]$ that maps ontic states to POVM elements defined on the $\psi$-space of the universe (but which act non-trivially only on the $\psi$-space of the system). Given a state $\ket{\psi}$ fixed for all time (we take it to be the $+1$ eigenstate for all $N$ qubits in the $Z$ basis, $\otimes^{N}_i \ket{z_{+}}_{i}$) and a measurement operator, quantum mechanics dictates that the probabilities are those given by the Born rule. Accordingly, we define the indicator function as: 
\eqn{
\xi_{M,m}[\lambda_S(t)] := \bra{\psi} \mathcal{O}_{M,m} [\lambda_S(t)] \ket{\psi},
} 
which gives the probabilities of obtaining the outcome $m$ when performing the measurement $M$ on a system in the ontic state $\lambda_S(t)$. If a measurement procedure $A$ is performed in region $\mathcal{A}$, independently of a measurement procedure $B$ in region $\mathcal{B}$, this defines a joint measurement procedure on the ontic state $\lambda_{\mathcal{AB}}$ in both regions, with outcomes $a,b$. The indicator function $\xi_{A,B,a,b}(\lambda_{\mathcal{AB}})$ of this joint measurement is associated with the function $\mathcal{O}_{A,a}[\lambda_{\mathcal{A}}] \mathcal{O}_{B,b}[\lambda_{\mathcal{B}}]$, which is just the product of the functions associated to the measurements in each local region. Note, however, that this does not imply that the joint indicator function is the product of the local indicator functions; $\xi_{A,B,a,b}(\lambda_{\mathcal{AB}}) \neq \xi_{A,a}(\lambda_{\mathcal{A} }) \xi_{B,b}(\lambda_{\mathcal{B} })$ in general. We will investigate this further below.

Putting it all together, the probability of obtaining outcome $m$ given the preparation $P$, transformation $T$ and measurement $M$ is given by (dropping the subscript $S$):
\eqn{
p(m|P,T,M) &=& \int \int d\lambda d\lambda' \, \mu(\lambda) \, \Gamma_T(\lambda', \lambda) \, \xi_{M,m}(\lambda')  \nonumber \\
&=& \int \int d\lambda d\lambda' \, \mu(\lambda) \, \Gamma_T(\lambda', \lambda) \, \bra{\psi} \mathcal{O}_{M,m} [\lambda'] \ket{\psi} 
\, , \nonumber \\
&&
}
where the integration is over the ontic state space $\Lambda_S$ of the system. By construction, this agrees with the probabilities predicted by quantum theory. Note that we can associate every indicator function with a unique POVM element, namely the operator given by $\mathcal{O}_{M,m}[\lambda]$ at $t=0$. Thus, the model is \tit{measurement non-contextual} as defined in Ref. \cite{SPE05}.

\tit{Locality---} We have seen that the DH-model is separable; this follows from the fact that the ontic state of a system is given by the union of the ontic states of its subsystems. We now show that it also satisfies dynamical locality. 

Consider the local operations on regions $\mathcal{A}, \mathcal{B}$, parameterised by settings $A,B$ and represented by the transition matrices $\Gamma_{A}(\lambda_{\mathcal{A}}, \lambda'_{\mathcal{A}} ), \, \Gamma_{B}(\lambda_{\mathcal{B}}, \lambda'_{\mathcal{B}} )$ respectively. We can ignore the local measurements, since by definition these do not change the ontic state. We consider deterministic transformations for clarity; the argument can easily be generalised to the non-deterministic case. We therefore associate the transition matrix $\Gamma_{A}(\lambda_{\mathcal{A}}, \lambda'_{\mathcal{A}} )$ with a unitary $\bar{U}_{A}$ parameterised by $A$, where $\bar{U}_{A}$ acts as the identity on all systems not in the region $\mathcal{A}$ (we define $\bar{U}_{B}$ symmetrically). This means that $\bar{U}_{A}$ commutes with all of the operators $\bar{X}_i,\bar{Y}_i,\bar{Z}_i$ in $\lambda_{\mathcal{B}}$, and vice versa. Hence:
\eqn{
\lambda'_{\mathcal{B}} &:=& \bar{U}^{\dagger}_{A} \bar{U}^{\dagger}_{B} \, \lambda_{\mathcal{B}} \, \bar{U}_{B} \bar{U}_{A} \nonumber \\
&=& \bar{U}^{\dagger}_{B} \, \lambda_{\mathcal{B}} \, \bar{U}_{B} \, ,
}
and the final state in region $\mathcal{B}$ depends explicitly only on $B$ and not on $A$. A symmetric argument holds for the final state $\lambda'_{\mathcal{A}}$, hence DL is satisfied. 

We have already remarked that the DH-model violates factorisation, and hence local causality, as is implied by Bell's theorem. To see this more explicitly, consider a state $\lambda_{\mathcal{AB}}$ at $t=0$ and a pair of local measurements $A,B$ on the sub-regions $\mathcal{A},\mathcal{B}$ resulting in outcomes $a,b$. These are represented in the DH-model by functions of the local ontic states, $\mathcal{O}_{A,a} [\lambda_{\mathcal{A}}]$ and $\mathcal{O}_{B,b} [\lambda_{\mathcal{B}}]$, with the corresponding joint measurement represented by the product:
\eqn{
\mathcal{O}_{A,B,a,b} [\lambda_{\mathcal{AB}}]  = \mathcal{O}_{A,a} [\lambda_{\mathcal{A}}] \mathcal{O}_{B,b} [\lambda_{\mathcal{B}}] \, .
}
Factorisation is equivalent to demanding that
\eqn{
 \xi_{A,B,a,b} (\lambda_{\mathcal{AB}}) = \xi_{A,a} (\lambda_{\mathcal{A} }) \xi_{B,b} (\lambda_{\mathcal{B}})
 }
holds for all ontic states. Given a state $\lambda_{\mathcal{AB}}$, we can represent an arbitrary ontic state by $U^{\dagger} \lambda_{\mathcal{AB}} U$ for some unitary $U$. Substituting this into the RHS and expanding, we obtain:
\eqn{
 \xi_{A,B,a,b} (\lambda_{\mathcal{AB}}) &=&  \bra{\psi} \, \mathcal{O}_{A,a} [U^{\dagger} \lambda_{\mathcal{A}} U] \, \mathcal{O}_{B,b} [U^{\dagger} \lambda_{\mathcal{B}} U] \,  \ket{\psi} \,  \nonumber \\
 &=&  \bra{\psi} \, (U^{\dagger} \bar{O}_{A,a} U ) \, (U^{\dagger} \bar{O}_{B,b} U ) \,  \ket{\psi} \,  \nonumber \\
 &=&  \bra{\psi} U^{\dagger} (\, \bar{O}_{A,a} \otimes \bar{O}_{B,b} \, )U \ket{\psi} 
  }
and it is clear that, for a general unitary acting on the states in both regions, the RHS will not factorise into a product of terms dependent on $(A,a)$ and $(B,b)$ respectively. 

\section{Discussion }

The failure of factorisation represents an obvious opportunity to challenge the DH-model's claims to locality. Indeed, if a violation of factorisation were as severe a transgression of `locality' as a violation of DL or S, there would be a problem. However, upon examination it is not at all obvious that factorisation has anything at all to do with locality. We have here presented a model in which physical systems can be assigned independent real physical states, which evolve locally in time and for which the measurement outcomes can depend only on the state of the systems upon which the measurement is performed. The critic is tasked with pointing out exactly what aspect of this picture is supposed to be `non-local' as a consequence of the non-factorisation of probabilities. 

Provided one does not interpret the lack of factorisability as an indication that the model is an incomplete description of reality (see below), it is hard to see how one would make this case. As Brown and Timpson \cite{BT14} have argued, the acceptance of intrinsic randomness in ontological models already undermines factorisation. 

Perhaps the most compelling problem with non-factorisation is that it represents a notion of ontological incompleteness. It is implicit that the ontic state is supposed to be a complete specification of all information that is in principle knowable about a physical system, including correlations between its parts. Without going so far as to presume that the ontic state determines the measurement outcomes (predetermination), the assumption of factorisation is a non-trivial constraint. In essence it says that \tit{it is not the measurement outcomes per se, but the correlations between them that are determined by the ontic state}. Thus, if I know the ontic state of $\lambda_{\mathcal{AB}}$, then a measurement on the part in $\mathcal{A}$ cannot provide any new information about the part in $\mathcal{B}$ beyond what is deducible from the measurement outcome $a$ and knowledge of the ontic state. If the probabilities did not factorise, then it would be possible to learn something new about $\mathcal{B}$ by doing measurements on $\mathcal{A}$, even though the entire state of reality $\lambda_{\mathcal{AB}}$ was already known. Where does this new information come from? Thus, as pointed out by Cavalcanti and Lal \cite{CAVL14}, approaches that abandon the assumption of factorisation do not constrain the possible correlations enough to be regarded as an explanation of them. However, we emphasise that this argument in favour of factorisation depends more on completeness than locality (see Butterfield \cite{BUT92} for a thorough elaboration of this point).

There is, perhaps, one more concern that could be raised about the model's supposed locality. The astute reader will have noticed that neither separability nor dynamical locality make any statements about probabilities, these being governed solely by the application of indicator functions. Since it is in the indicator functions where the non-factorisation occurs, we might wonder whether all of the interesting physics has simply been hidden in the abstract form of these functions. Perhaps the model only seems local, but in fact the non-factorisation of probabilities is concealing some form of non-locality, and the DH-model represents little more than a sleight-of-hand in shifting the non-locality to these functions.

However, recall that the indicator functions are subject to a rather non-trivial locality constraint, namely, the indicator function associated to a measurement procedure within a given region of space-time can only depend on the ontic states in that region. This is why, as noted in the Background, the purely ontological constraint of dynamical locality is sufficient to imply signal locality at the level of probabilities. But suppose we drop the requirement that the model's predictions should agree with quantum mechanics and consider the set of all generalised probabilistic models in which the kinematics and dynamics are restricted to be separable and dynamically local, but where the indicator functions can now be arbitrary functions of the local ontic state (as defined in the DH-model). We can then ask: do these models imply any constraints on the probabilities that are stronger than just signal locality? If so, this would lend even more support to the case that kinematical and dynamical locality represent a non-trivial constraint on ontological models, that is, constraints which cannot be ``undone" by toying with the indicator functions. Generalized probabilistic theories based on the DH-model may be interesting to study in their own right, particularly if there are interesting generalisations that are amenable to experimental tests. 

\tit{Criticisms---} We have presented the DH-model primarily to prove a point: that quantum mechanics is compatible with an ontological model that is both separable and dynamically local. But is the DH-model a mere contrivance, or an acceptable picture of reality?

There are a number of difficulties to be addressed if one is to take the model seriously. First, we mentioned the `excess baggage' problem, which requires the ontic state space of a system to be rather absurdly large. Previously, this was known to be a feature of $\psi$-ontic models \cite{LEI14}, so it is interesting that it resurfaces in a setting where $\psi$ has no part in the ontic state. The equivalence between the Schr\"{o}dinger and Heisenberg pictures leads one to speculate that there might exist a $\psi$-ontic model that is equivalent to the DH-model (and therefore also separable), in which case the `excess baggage' result would follow. Otherwise, it remains an open question whether the model can be cured of this problem. 

There is also the issue of under-determination. We have noted that the ontic state of a physical system may differ depending on how it is prepared and transformed (eg. depending on how the dynamics is purified to ancilla qubits), and this is true even if both preparations would result in the same reduced density matrix under the standard formalism. Thus, a quantum system described by a known density matrix may be in any one of a continuous family of ontic states, and no local measurement on the system can distinguish which state it is in. Timpson \cite{TIM03} and Wallace and Timpson \cite{WALT05} have argued that, by analogy with the unphysical gauge freedom of classical electrodynamics, such operationally indistinguishable states should correspond to a single state of reality, contra the DH-model. Deutsch \cite{DEUVIN} has countered that states arising from different preparation procedures, even if they have the same density matrix, \tit{can} be operationally distinguished if one is allowed to include measurements on the state at other times, e.g. during the process of its preparation. We have seen that in the DH-model, this under-determination is just an expression of the fact that the model is preparation and transformation contextual, that is, the ontic states are determined by details of the experimental procedure that are not relevant to the density matrix of the system. Seen in this light, the under-determinism identified by Wallace and Timpson is only a problem to the extent that contextuality is a problem, but \tit{any} model that reproduces quantum mechanics must be at least preparation contextual \cite{SPE05}. We note that, since classical electromagnetism is arguably non-contextual, its gauge freedom is significantly more problematic, because the choice of gauge cannot even be ascertained from the contextual details of the experimental procedure. The DH-model therefore seems to fare better against the charge of underdeterminism.

Finally, no discussion of Deutsch and Hayden's work would be complete without mentioning information flow, since the model was designed for the express purpose of elucidating this concept. As originally conceived by the authors, their formalism allows one to explicitly trace the path of quantum information using the ontic states as physical carriers. In this way it can be shown that information is always localised to systems in a region of space at a given time, which follow time-like geodesics that do not violate causality, and this is true even in experiments such as the Bell scenario and quantum teleportation (see also Ref. \cite{HORS} for a treatment of entanglement swapping). The DH-model presented here retains all of these features; one can trace the information flow as usual by following the dependence of the ontic states on the information-bearing parameters, and since the ontic states follow time-like paths in space-time up to the point of measurement, so does the information. It was noticed by Deutsch and Hayden that their model implied a concept of information that is `locally inaccessible': information that appears explicitly in the ontic state of a system, but which cannot be revealed by local measurements on that system alone. To recover the information, a joint measurement on the system and the systems with which it is entangled is required. We note that, if all information were locally accessible, the ontic state would be deducible purely from local measurements and the model would be factorisable (contrary to quantum mechanics). Hence in the DH-model, locally inaccessible information is a manifestation of non-factorisation. 

\section{Conclusions and outlook}

We have shown that, perhaps contrary to expectations, it is possible to formulate an ontological model that is separable and dynamically local. We based the model on Deutsch and Hayden's formalism but without relying on the many-worlds interpretation, thereby making it susceptible to Bell's theorem. Applying the latter, we found that the model nevertheless satisfies both \tit{separability} and \tit{locality}, although it does not satisfy \tit{factorisation}. Nevertheless, we argued that the model is as local as possible while still agreeing with the predictions of quantum mechanics.

If one were to take the Deutsch-Hayden model seriously, one would still have to outline exactly what the physical qubits represent and how they are connected to space-time. The manifest separability and locality of the model might make it a useful platform for building toy models beyond quantum mechanics, particularly where relativity and gravitation are concerned. 

Within the setting of generalized probabilistic theories, the Deutsch-Hayden model suggests an intriguing new approach based on `algebra-valued reality'. Supposing that the elements of reality are represented by algebraic objects whose behaviour satisfies certain `locality' requirements, such as separability and dynamical locality, we speculate that this places non-trivial constraints on generalised probabilistic theories. This might provide a way to defend against the charge of incompleteness, if one can show that correlations are constrained by the algebra in a meaningful way, despite being non-factorisable. We leave this to future work.

\bibliography{DHref}{}
\bibliographystyle{unsrt}

\end{document}